\documentclass[prl,aps,twocolumn,structural,superscriptaddress,amsmath,amssymb,showpacs]{revtex4}
\usepackage[dvips]{graphicx}
\usepackage{color}
\usepackage{amssymb}
\usepackage{amsmath}
\usepackage{units}
\usepackage[latin1]{inputenc}

\begin{document}

\title
{Improvement of structural, electronic, and magnetic properties of
Co$_2$MnSi thin films by He$^+$-irradiation}

\author{O.~Gaier}
\affiliation{Fachbereich Physik and Forschungszentrum OPTIMAS,
Technische Universit\"at Kaiserslautern,
Erwin-Schr\"odinger-Stra\ss e 56, D-67663 Kaiserslautern, Germany}

\author{J.~Hamrle}
\affiliation{Fachbereich Physik and Forschungszentrum OPTIMAS,
Technische Universit\"at Kaiserslautern,
Erwin-Schr\"odinger-Stra\ss e 56, D-67663 Kaiserslautern, Germany}

\author{B.~Hillebrands}
\affiliation{Fachbereich Physik and Forschungszentrum OPTIMAS,
Technische Universit\"at Kaiserslautern,
Erwin-Schr\"odinger-Stra\ss e 56, D-67663 Kaiserslautern, Germany}

\author{M.~Kallmayer}
\affiliation{Institut f\"ur Physik, Johannes
Gutenberg-Universit\"at, D-55099 Mainz, Germany}

\author{P.~P\"orsch}
\affiliation{Institut f\"ur Physik, Johannes
Gutenberg-Universit\"at, D-55099 Mainz, Germany}

\author{G.~Sch\"onhense}
\affiliation{Institut f\"ur Physik, Johannes
Gutenberg-Universit\"at, D-55099 Mainz, Germany}

\author{H.~J.~Elmers}
\affiliation{Institut f\"ur Physik, Johannes
Gutenberg-Universit\"at, D-55099 Mainz, Germany}

\author{J.~Fassbender}
\affiliation{Forschungszentrum Dresden-Rossendorf e.V., Institut
f\"ur Ionenstrahlphysik und Materialforschung, Bautzner
Landstrasse 128, D-01328 Dresden, Germany}

\author{A.~Gloskovskii}
\affiliation{Institute of Inorganic and Analytical Chemistry,
Johannes Gutenberg-Universit\"at, D-55099 Mainz, Germany}

\author{C.~A.~Jenkins}
\affiliation{Institute of Inorganic and Analytical Chemistry,
Johannes Gutenberg-Universit\"at, D-55099 Mainz, Germany}

\author{G.~H.~Fecher}
\affiliation{Institute of Inorganic and Analytical Chemistry,
Johannes Gutenberg-Universit\"at, D-55099 Mainz, Germany}

\author{C.~Felser}
\affiliation{Institute of Inorganic and Analytical Chemistry,
Johannes Gutenberg-Universit\"at, D-55099 Mainz, Germany}

\author{E.~Ikenaga}
\affiliation{SPring-8 JASRI, Hyogo, 679-5198, Japan}

\author{Y.~Sakuraba}
\affiliation{Magnetic Materials Laboratory, Institute for
Materials Research (IMR), Tohoku University, Katahira 2-1-1,
Sendai 980-8577, Japan}

\author{S.~Tsunegi}
\affiliation{Department of Applied Physics, Graduate School of
Engineering, Tohoku University, Aoba-yama 6-6-05, Sendai 980-8579,
Japan}

\author{M.~Oogane}
\affiliation{Department of Applied Physics, Graduate School of
Engineering, Tohoku University, Aoba-yama 6-6-05, Sendai 980-8579,
Japan}

\author{Y.~Ando}
\affiliation{Department of Applied Physics, Graduate School of
Engineering, Tohoku University, Aoba-yama 6-6-05, Sendai 980-8579,
Japan}

\date{\today}

\begin{abstract}

The influence of \unit[30]{keV} He$^+$ ion irradiation on
structural, electronic and magnetic properties of Co$_2$MnSi thin
films with B2 order was investigated. It was found, that
irradiation with light ions can improve the local chemical order.
This provokes changes of the electronic structure and
element-specific magnetization towards the bulk properties of the
well-ordered Co$_2$MnSi Heusler compound with L2$_1$ structure.
\end{abstract}


\maketitle%
Half-metallic ferromagnetism, theoretically expected for the
L2$_1$ ordered Co$_2$MnSi Heusler compound \cite{gal02}, in
combination with a large band gap of \unit[0.4]{eV} \cite{fuj90}
to \unit[0.8]{eV} \cite{pic02} for the minority spin states and a
high Curie temperature of \unit[985]{K} \cite{bro00} makes this
Heusler compound a very promising candidate for the modern field
of spintronic devices. In recent years, great progress has been
made in the fabrication of magnetic tunnel junctions (MTJ) with a
single integrated Co$_2$MnSi electrode \cite{sch04, sak05, ish06,
sak06-apl88, sak06-apl89}, and an exceptionally high TMR effect of
\unit[570]{\%} at \unit[2]{K} has been reported for a MTJ
structure with both electrodes consisting of Co$_2$MnSi
\cite{sak06-apl88}. In the latter case, the spin polarization
estimated by Julli\`{e}re's formula was reported to be
\unit[89]{\%} and \unit[83]{\%} for the bottom and top electrode,
respectively. At room temperature (RT), however, the TMR effect is
largely reduced. Moreover, spin polarization, experimentally
observed on single crystalline Co$_2$MnSi films at RT, remains
with \unit[12]{\%} \cite{wan05} far below the theoretically
predicted \unit[100]{\%}. Strongly supported by {\itshape ab
initio} calculations \cite{pic04, ger07}, partial chemical
disorder is assumed to be one of the possible reasons for this
discrepancy.

Annealing at high temperatures, typically in the range of
\unit[400-500]{°C}, is a conventional way to reduce the chemical
disorder in the deposited Co$_2$MnSi films. However, high
temperature annealing often leads to interdiffusion and local
changes of the stoichiometry in Heusler compounds \cite{ber06}.
For FePt thin layers, however, it has been demonstrated that the
degree of chemical order can alternatively be controlled by
post-growth irradiation with He$^+$ ions \cite{rav00, ber03}. In
both completely disordered and partially L1$_0$ ordered FePt films
an enhancement of long range order was found after \unit[130]{keV}
He$^+$ ion irradiation at moderate processing temperatures. The
post-growth irradiation process thus improves local order leaving
the large-scale elemental distribution intact. Moreover, the
initial crystallographic structure is maintained due to the
absence of extended collision cascades. In view of the successful
results for FePt and other binary systems \cite{fas04, fas08}, the
question arises whether the light-ion irradiation technique is
also applicable for the improvement of chemical order in
Co$_2$MnSi, representing, as a ternary compound, a more complex
system.

In this work, we investigate the effect of \unit[30]{keV} He$^+$
ion irradiation on chemical order of Co$_2$MnSi thin films. The
samples are exposed to different fluences of He$^+$ ions. The
information about the ordering properties is obtained from X-ray
diffraction (XRD), X-ray absorption and circular magnetic
dichroism (XAS/XMCD), and photoemission spectroscopy at high
energies (HAXPES).


A (001)-oriented Co$_2$MnSi layer of \unit[30]{nm} thickness was
grown on a Cr-buffered MgO(001) substrate by means of inductively
coupled plasma (ICP) assisted magnetron sputtering. The chemical
composition of the deposited film was nearly stoichiometric (Co:
48.9~$\%$; Mn: 24.7~$\%$; Si: 26.4~$\%$). Annealing at
\unit[350]{°C} followed the deposition of the Co$_2$MnSi layer.
Subsequently, a \unit[1.3]{nm} Al capping layer was deposited to
prevent oxidation of the Co$_2$MnSi film. To ensure equal initial
conditions for all irradiation experiments a single
\unit[1]{in$^2$} Co$_2$MnSi sample was prepared which was cut into
\unit[5$\times$5]{mm$^2$} pieces before carrying out the
irradiation. The entire surface of the \unit[5$\times$5]{mm$^2$}
Co$_2$MnSi samples has been irradiated with \unit[30]{keV} He$^+$
ions at ambient temperature using a DANFYSIK low energy ion
implanter. The ion fluence was varied between \unit[10$^{14}$] and
\unit[10$^{16}$]{ions/cm$^2$} using a beam current between
\unit[0.5]{} and \unit[2.5]{$\mu$A/cm$^2$}.

XAS/XMCD measurements were performed at the beamline UE56/1-SGM at
BESSY II in Berlin. The samples were magnetically saturated by an
external magnetic field of \unit[1.6]{T} applied perpendicular to
the film surface. The x-ray absorption was measured in
transmission giving access to magnetic moments averaged along the
film normal~\cite{kal06}. The element-specific magnetic moments
per 3$d$-hole were determined by a sum rule analysis. The HAXPES
experiments were performed at the beamline BL47XU of SPring-8
(Japan). The photon energy was fixed at 7.940~keV. The kinetic
energy of the photoemitted electrons was analyzed by means of a
hemispherical analyzer (Scienta R4000-12kV) with an overall energy
resolution of 250~meV~\cite{bal06}.



\begin{figure}
\includegraphics[width=0.95\columnwidth]{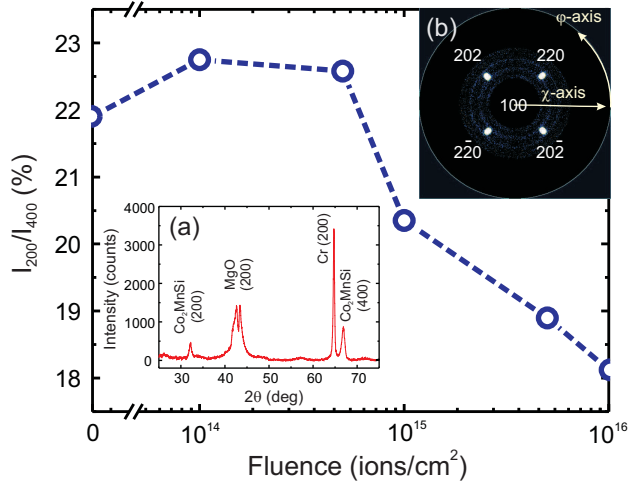}
\caption{%
\label{figure1}%
(Color online) Ratio of the (200) and (400) integrated XRD
intensities in dependence of the applied ion fluence. An increase
at \unit[1$\times$10$^{14}$]{} and
\unit[5$\times$10$^{14}$]{ions/cm$^2$} is clearly visible. The
insets show (a) the $\theta$-2$\theta$-scan of the as prepared
MgO/Cr(\unit[40]{nm})/Co$_2$MnSi(\unit[30]{nm})/Al(\unit[1.3]{nm})
film and (b) the corresponding pole figure of the fundamental
(220) reflections. }
\end{figure}

\begin{figure}
\includegraphics[width=0.7\columnwidth]{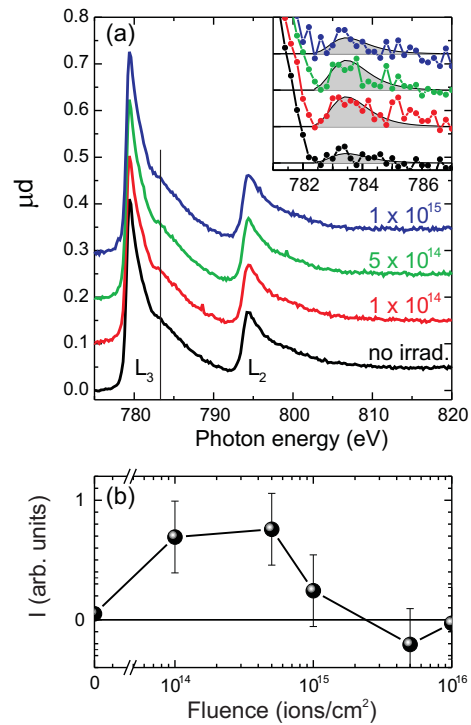}
\caption{%
\label{figure2}%
(Color online) (a) Transmission XAS spectra at the Co L$_{2,3}$
absorption edge from Co$_2$MnSi films irradiated with
\unit[30]{keV} He$^+$ ions at the indicated fluences. The inset
shows the energy range of the Co satellite peak after linear
background subtraction together with a fit of the peak area. (b)
Dependence of the intensity of the satellite peak on the He$^+$
ion irradiation. }
\end{figure}

Results of XRD structural characterization of the as deposited
Co$_2$MnSi film are shown in the insets of Fig.~\ref{figure1}. The
(220) equivalent reflections were observed with fourfold symmetry
in the x-ray pole figure scan providing evidence of the epitaxial
growth of the Co$_2$MnSi layer. The x-ray $\theta$-2$\theta$
diffraction pattern exhibits clear (200) B2 superstructure
reflections. Since no (111) reflections, indicative for the L2$_1$
phase, were detected in the pole figure scan (not shown here), it
is concluded that the fabricated film was predominantly of B2
order. The presence of a certain amount of A2 type disorder cannot
be excluded from these results.

After the ion bombardment, the (111) equivalent reflections are
still absent in the pole figure scans while (220) pole figures
remain similar to the one presented in the inset of
Fig.~\ref{figure1}. A detailed analysis of
$\theta$-2$\theta$-scans recorded from irradiated samples reveals
an increase of the ratio of (200) and (400) integrated intensities
for the fluences of \unit[1$\times$10$^{14}$]{} and
\unit[5$\times$10$^{14}$]{ions/cm$^2$} (see Fig~\ref{figure1}).
Comparing with simulated peak intensities this result clearly
suggests a qualitative improvement of the B2 order in the
irradiated Co$_2$MnSi films. A transition to the L2$_1$ ordered
phase, however, was not observed.


Figure~\ref{figure2} shows XAS data recorded at the Co $L_{2,3}$
edge comparing irradiated and non-irradiated samples. A satellite
peak appears at 3.8~eV above the $L_3$ absorption edge. This
feature has been reported for thin Co$_2$MnSi films by different
groups \cite{sch04, tel06, sai07}. In particular, it has been
demonstrated that the intensity of the satellite directly
correlates with the degree of ordering inside the Co$_2$MnSi layer
\cite{tel06}. The intensity of the Co L$_3$ satellite is plotted
in Fig.~\ref{figure2}(b) as a function of ion fluence. One can
clearly see that at fluences of \unit[1$\times$10$^{14}$] and
\unit[5$\times$10$^{14}$]{ions/cm$^2$} the intensity of the
satellite peak increases with respect to the case of the
non-irradiated sample [see also inset in Fig.~\ref{figure2}(a)].
This suggests, in good agreement with the results of XRD
characterization, that in this particular range of applied
fluences the local order in the bulk of Co$_2$MnSi films is
increased after the irradiation. For higher fluences, however, the
intensity of the satellite decreases indicating an increasing
level of disorder introduced by the ion bombardment beyond
\unit[5$\times$10$^{14}$]{ions/cm$^2$}.

\begin{figure}
\includegraphics[width=0.7\columnwidth]{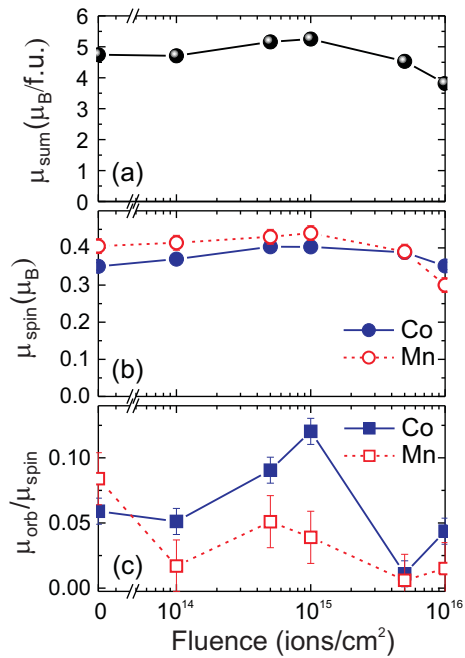}
\caption{%
\label{figure3}%
(Color online) (a) Saturation magnetization of Co$_2$MnSi films
irradiated with \unit[30]{keV} He$^+$ ions at different fluences
obtained from XMCD measurements. (b) Element-specific spin
magnetic moments per 3d-hole, and (c) ratios of orbital and spin
magnetic moments. }
\end{figure}
\begin{figure}
\includegraphics[width=0.8\columnwidth]{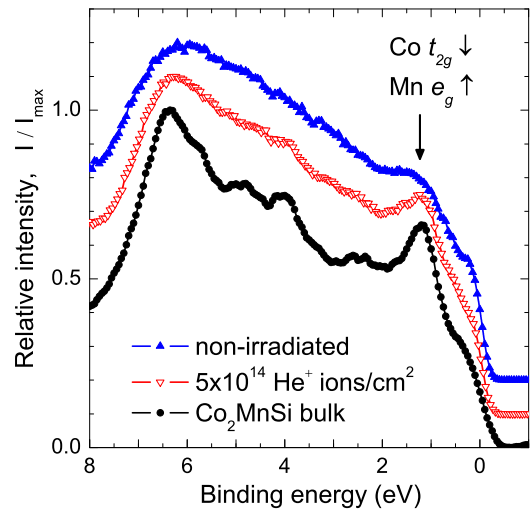}
\caption{%
\label{figure4}%
(Color online)
HAXPES ($h\nu=7.94$~keV) spectra of Co$_2$MnSi films
irradiated with \unit[30]{keV} He$^+$ ions in comparison to a
non-irradiated sample and a cleaved Co$_2$MnSi bulk sample.
The enhanced intensity of Co and Mn $d$-states at about 1.3~eV,
characteristic for bulk Co$_2$MnSi, is marked by an arrow.
}
\end{figure}

We determined the element-specific magnetic moments per 3$d$-hole
of the Co$_2$MnSi films exposed to different ion fluences. The
corresponding values are shown in Fig.~\ref{figure3}. The volume
magnetic moments of both Co and Mn are in good agreement with
values reported in Ref.~\cite{tel06} for the same kind of
Co$_2$MnSi samples. As a result of ion irradiation we observe an
increase of the spin magnetic moments on both Co and Mn atoms up
to a fluence of \unit[1$\times$10$^{15}$]{ions/cm$^2$} and a
subsequent decrease for higher fluences (Fig.~\ref{figure3}(b)).
This is again a strong hint of the improvement of chemical order
inside the Co$_2$MnSi layer introduced by the irradiation with
He$^+$ ions. The ratio of orbital and spin magnetic moments
(Fig.~\ref{figure3}(c)) shows for Co a pronounced peak at a
fluence of \unit[1$\times$10$^{15}$]{ions/cm$^2$} indicating a
reduction of symmetry. At the Mn site the effect is less
pronounced.

Assuming the number of 3d-holes for Co and Mn to be 2.4 and 4.5
per atom \cite{sch04}, respectively, and taking into account an
additional correction factor of 1.5 for the number of Mn 3d-holes
\cite{dur97}, which is necessary due to the mixing of two Mn $j$
levels, the saturation magnetization of the investigated
Co$_2$MnSi films is calculated. The corresponding values are
presented in Fig.~\ref{figure3}(a). The saturation magnetization
increases up to a fluence of
\unit[1$\times$10$^{15}$]{ions/cm$^2$} reaching the value of
\unit[5]{$\mu_{B}$/f.u.} theoretically predicted for the
well-ordered Co$_2$MnSi bulk and decreases for fluences beyond
\unit[1$\times$10$^{15}$]{ions/cm$^2$}. This behavior is confirmed
by the results of SQUID measurements (not shown here) and provides
a further hint for the improvement of chemical order provoked
inside the Co$_2$MnSi layer by the irradiation with He$^+$ ions.

The HAXPES valence band spectra are shown in Fig.~\ref{figure4},
comparing the non-irradiated film and the bulk reference sample
with the sample irradiated with the optimum fluence of
\unit[5$\times$10$^{14}$]{ions/cm$^2$}. The inelastic mean free
path of the 8~keV electrons is expected to be about 13~nm in
AlO$_x$ and 8~nm in Co$_2$MnSi~\cite{tan93}. This allows for an
investigation of the bulk electronic properties of Co$_2$MnSi
films below the capping layer~\cite{fec08}.

The valence band spectrum of the non-irradiated film has a wide
maximum in the energy range from about 7~eV to 0~eV without
distinct features. In particular, the pronounced peak of the
$d$-states - being well resolved in the valence spectrum of the
bulk sample - is largely smeared out. This broadening points to a
high disorder in the sample. After irradiation, the valence band
spectrum of the thin film resembles much closer that of the bulk
material, in particular close to the Fermi energy. The peak at
about 1.3~eV is due to emission from flat $d$-bands that belong to
minority $t_{2g}$ states localized in the Co planes as well as
highly localized Mn $d$-majority $e_g$ states~\cite{bal06,fec08}.
The similarity of the electronic structure of the irradiated
sample and the bulk sample provides clear evidence on the
structural improvement of the Co$_2$MnSi film after irradiation.


In conclusion, motivated by a successful application of light-ion
irradiation for the improvement of chemical order in thin FePt
layers, the effect of a \unit[30]{keV} He$^+$ ion bombardment on
structural, electronic and magnetic properties of Co$_2$MnSi thin
films was investigated. The results of XRD, XMCD and HAXPES
investigations demonstrate that the light-ion irradiation
technique has the potential to invoke local chemical order in
Co$_2$MnSi Heusler films without the need of high-temperature
annealing. The presence of local chemical order in Heusler
compounds is absolutely crucial for the appearance of
half-metallic ferromagnetism and the desired complete
spin-polarization of conduction electrons. The standard procedure
of high-temperature annealing close to the melting point needed to
invoke the chemical order, that is applied to bulk materials, is
inhibited for the preparation of thin film multilayer structures
because the interdiffusion of different layers will destroy the
stoichiometric composition. Light-ion irradiation provides a
formidable method to overcome this problem. Future studies will
show whether a combination of mild annealing and irradiation
further improves the local order. Our results also implicate that
light-ion irradiation can improve local chemical order in other
ordered compounds that are related to the Co$_2$MnSi Heusler
compound, i.e. in other half-metallic Co$_2$YZ (Y=3d metal, Z=main
group element) Heusler compounds and in Ni$_2$MnGa shape memory
compounds.


The project was financially supported by the Research Unit 559
\emph{"New materials with high spin polarization"} funded by the
Deutsche Forschungsgemeinschaft and by the NEDO International
Joint Research Grant Programm 2004/T093. M.~K. and H.~J.~E. wish
to thank S.~Cramm for support at the beamline and the
BMBF(ES3XBA/5) for a total expenses grant. We also thank
H.~Schneider for SQUID measurements, B.~Balke for producing the
bulk sample and K.~Kobayashi and the staff of SPring-8 for their
help during the beamtime.

\end{document}